\documentstyle[preprint,aps,prc,floats,epsfig]{revtex}
\tighten
\begin{document}
\draft
\preprint{}

\preprint{\parbox{2.0in}{\noindent AMES-HET-97-5}}

\title{\ \\[5mm] Flavor-changing top quark decays in R-parity-violating
                 supersymmetric models}

\author{\ \\[2mm] Jin Min Yang$^{1,2}${\footnote{
                                Address after March of 2000: 
   Institute of Theoretical Physics, Academia Sinica, Beijing, China}}, 
  Bing-Lin Young$^1$,  X. Zhang$^{3,4}$ }

\address{\ \\[2mm]   
$^1$ {\it  Department of Physics and Astronomy, Iowa State University,\\
               Ames, Iowa 50011}\\[2mm]
$^2$ {\it International Institute of Theoretical and Applied Physics,\\
          Iowa State University, Ames, Iowa 50011}\\[2mm]
$^3$ {\it CCAST (World Laboratory), P.O.Box 8730, Beijing 100080, China}\\[2mm]
$^4$ {\it Institute of High Energy Physics, Academia Sinica, \\
         Beijing 100039, China } }

\maketitle

\begin{abstract}
The flavor changing top quark decays $t\rightarrow cV $ ($V=Z,\gamma, g$)
induced by R-parity-violating couplings in the minimal 
supersymmetric standard model (MSSM) are evaluated.
We find that the decays $t \rightarrow cV$ can be  significantly enhanced 
relative to those in the R-parity conserving SUSY model. Our results show 
that the top quark FCNC decay can be as large as
$Br(t \rightarrow cg) \sim 10^{-3}$,
$Br(t \rightarrow cZ) \sim 10^{-4}$ and $Br(t \rightarrow c\gamma)
\sim 10^{-5}$, which may be observable at the upgraded Tevatron and/or
the LHC.
\end{abstract}



The unexpected large mass of the top quark suggest that it may be more 
sensitive to new physics than other fermions.
In the standard model (SM) the flavor changing neutral current (FCNC)
decays of the top quark, $t\rightarrow cV$, suppressed by GIM, are found 
to be far below the detectable level [1,2].
So, searching for FCNC top decays serves as a powerful probe to 
effects of new physics. The CDF [3,4] and D0 [5] collaborations 
have reported interesting bounds on these decays [4].
Undoubtedly more stringent bounds will be obtained in the 
future at the Tevatron upgrade and the LHC. 

Systematic theoretical study of the experimental
observability for FCNC top quark decays at the Tevatron and the
LHC has been made in [6,7]. The results show that
the detection sensitivity can be significant [6,7]:
\begin{eqnarray} \label{level1}
Br(t \rightarrow cZ)&      \simeq & 4\times 10^{-3} (6\times 10^{-4}),\\ 
\label{level2}
Br(t \rightarrow c\gamma) &\simeq & 4\times 10^{-4}(8\times 10^{-5}),\\
\label{level3}
Br(t \rightarrow cg)     & \simeq & 5\times 10^{-3}(1\times 10^{-3}),
\end{eqnarray}
at the upgraded Tevatron of integrated luminosity of 10 (100) fb$^{-1}$.
The two electroweak modes can be improved several fold  
at the LHC with similar integrated luminosities:
\begin{eqnarray} 
Br(t \rightarrow cZ)     & \simeq & 8\times 10^{-4} (2\times 10^{-4}),\\
 \label{level5}
Br(t \rightarrow c\gamma)& \simeq & 2\times 10^{-5}(5\times 10^{-6}).
\end{eqnarray}

Despite the above interesting experimental possibilities,  
there is no demonstration in  the 
minimal supersymmetric model (MSSM) which is the most favored
candidate for physics beyond the standard model, that such limits can be
 realized. In MSSM conserving R-parity, the 
predictions for branching ratios of these FCNC top quark 
decays were found to be significantly below the above detectable 
levels[8]. In this paper we will show that in the case of R-parity 
violating MSSM [9, 10] with the existing 
bounds on the R-parity violating couplings that violate the baryon 
number,  $Br(t \rightarrow cV)$
might reach the detectable level at the upgraded Tevatron and the LHC. 
However, as shown below, 
the effects of the lepton number violating 
 $\lambda'$ couplings in FCNC top decays
are negligibly small under the current constraints.   

 In MSSM the superpotential with
 R-parity violating is given by [10]
\begin{equation} 
{\cal W}_{\not \! R}=\frac{1}{2}\lambda_{ijk}L_iL_jE_k^c
+\lambda_{ijk}^{\prime}L_iQ_jD_k^c
+\frac{1}{2}\lambda_{ijk}^{\prime\prime}\epsilon^{abd}U_{ia}^cD_{jb}^cD_{kd}^c
+\mu_iL_iH_2,
\end{equation}
where $L_i(Q_i)$ and $E_i(U_i,D_i)$ are the left-handed
lepton (quark) doublet and right-handed lepton (quark) singlet chiral 
superfields. $i,j,k$ are generation indices and $c$ denotes charge conjugation. 
$a$, $b$ and $d$ are the color indices and $\epsilon^{abd}$ is the total 
antisymmetric tensor.
$H_{1,2}$ are the Higgs-doublets chiral superfields.
The  $\lambda_{ijk}$ and $\lambda^{\prime}_{ijk}$ are lepton-number 
violating (${\large \not} \! L$) couplings, $\lambda^{\prime\prime}_{ijk}$ baryon-number 
violating (${\large \not} \! B$) couplings. 
Constraints on these couplings have been  obtained 
from various low energy processes [11-20] and their phenomenologies 
at hadron and lepton colliders have also been investigated recently 
by a number of authors [19]. 

 Although it is theoretically possible to have both ${\large \not} \! B$
and ${\large \not} \! L$ interactions, the non-observation
of proton decay prohibits their simultaneous
presence [14]. We, therefore, assume the existence of either 
${\large \not} \! L$
couplings or ${\large \not} \! B$ couplings, and investigate  their separate
effects in top quark decays.

The FCNC decays  $t\rightarrow cV$ can be
induced by either the $\lambda^{\prime}$ or $\lambda^{\prime\prime}$
coupling at the one loop level.
In terms of the four-component Dirac notation, the Lagrangian of the
${\large \not} \! L$  couplings $\lambda^{\prime}$ and ${\large \not} \! B$
 couplings $\lambda^{\prime\prime}$ are given by
\begin{eqnarray}
{\cal L}_{\lambda^{\prime}}&=&-\lambda^{\prime}_{ijk}
\left [\tilde \nu^i_L\bar d^k_R d^j_L+\tilde d^j_L\bar d^k_R\nu^i_L
       +(\tilde d^k_R)^*(\bar \nu^i_L)^c d^j_L\right.\nonumber\\
& &\hspace{1cm} \left. -\tilde e^i_L\bar d^k_R u^j_L
       -\tilde u^j_L\bar d^k_R e^i_L
       -(\tilde d^k_R)^*(\bar e^i_L)^c u^j_L\right ]+h.c.,\\
{\cal L}_{\lambda^{\prime\prime}}&=&-\frac{1}{2}\lambda^{\prime\prime}_{ijk}
\left [\tilde d^k_R(\bar u^i_R)^c d^j_R+\tilde d^j_R(\bar d^k_R)^c u^i_R
       +\tilde u^i_R(\bar d^j_R)^c d^k_R\right ]+h.c.
\end{eqnarray}
where the color indices in ${\cal L}_{\lambda^{\prime\prime}}$ are 
totally antisymmetric as in (6). 
\vspace{0.5cm}

Let us first consider
$t\rightarrow cV$ induced by ${\large \not} \! L$ couplings.
The relevant Feynman diagrams are shown in Fig.1.
\begin{figure}[tb]
\begin{center}
\psfig{figure=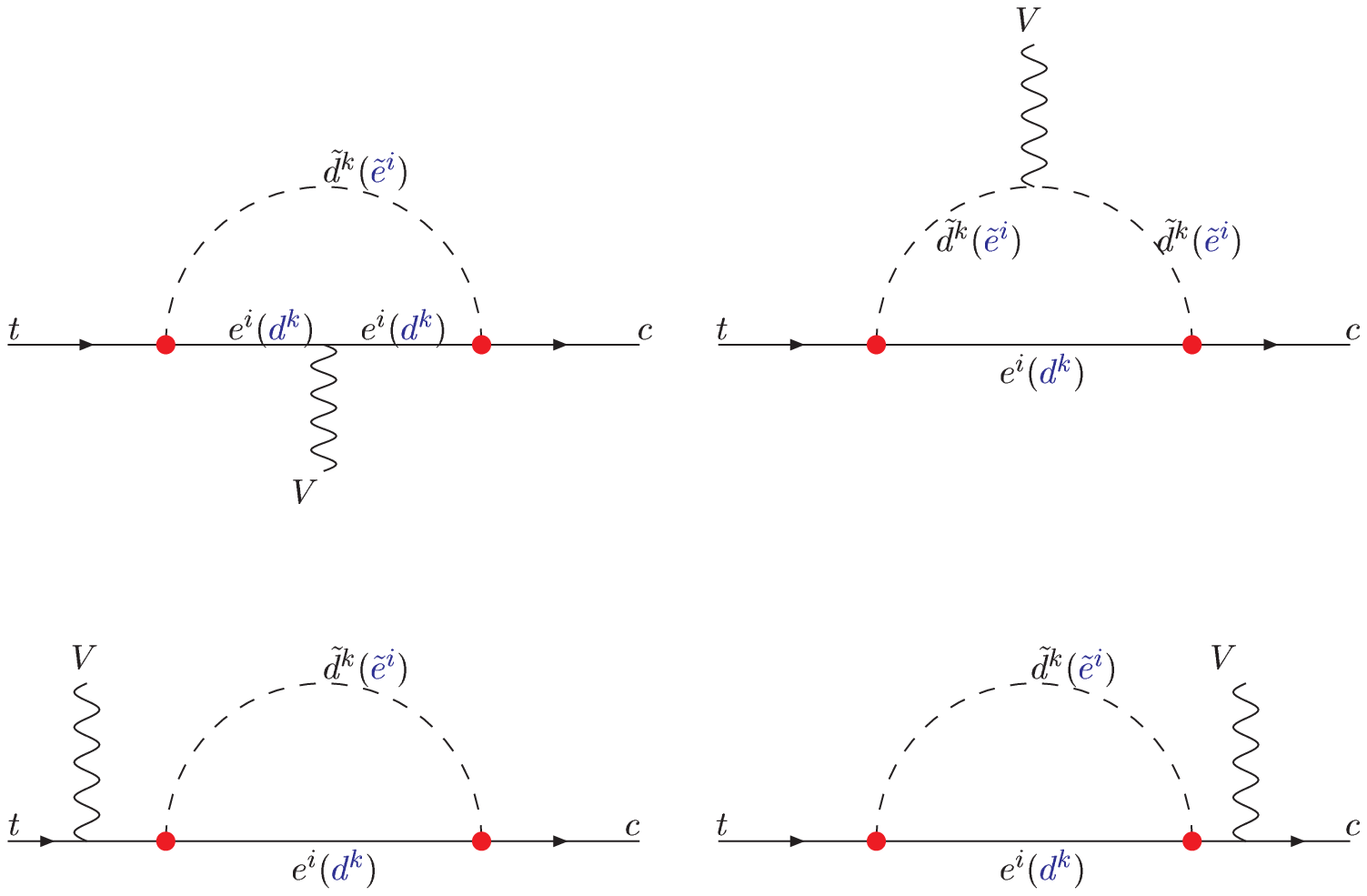,width=350pt,angle=0}
\end{center}
\vspace*{-0.5cm}
\caption{ Feynman diagrams for $t\rightarrow cV $ ($V=Z,\gamma, g$
for quarks and squarks; $V=Z,\gamma$ for leptons and sleptons) 
induced by L-violating couplings. The blobs denote L-violating vertex.} 
\label{fig1}
\end{figure}
At one-loop level, they give rise to effective $tcV$ vertices of the form
\begin{eqnarray} \label{verz}
V^{\mu}(tcZ)&=&ie\left[\gamma^{\mu}P_LA^Z
              +ik_{\nu}\sigma^{\mu\nu}P_R B^Z \right],\\
V^{\mu}(tc\gamma)&=&ie\left[ ik_{\nu}\sigma^{\mu\nu}P_R B^{\gamma}\right ],\\
\label{verg}
V^{\mu}(tcg)&=&ig_sT^a\left[ ik_{\nu}\sigma^{\mu\nu}P_R B^g \right ],
\end{eqnarray}
where  $P_{R,L}=\frac{1}{2}(1\pm \gamma_5)$ and $k$ is the momentum 
of the vector boson. The form factors $A^Z$, $B^Z$, etc., are obtained by
identifying $A^Z=A^Z_1+A^Z_2$ and $B^V=B^V_1+B^V_2,(V=Z,\gamma,g)$, where
$A^{\gamma,g}$ are found to be zero due to the gauge invariance while others
are given by 
\begin{eqnarray}
A^Z_1&=&\frac{1}{16\pi^2}\lambda^{\prime}_{i2k}\lambda^{\prime}_{i3k}
\left \{ (v_c+a_c)B_1(M_t,M_{e^i},M_{\tilde d^k})\right.\nonumber\\
& & - (v_e+a_e)\left [2c_{24}-\frac{1}{2}+M_Z^2(c_{12}+c_{23})\right]
                  (-p_t,p_c,M_{e^i},M_{\tilde d^k},M_{e^i})\nonumber\\
& &\left. + \xi_V\left[2c_{24}
+M_t^2\left(c_{11}-c_{12}+c_{21}-c_{23}\right)\right]
                  (-p_t,k,M_{e^i},M_{\tilde d^k},M_{\tilde d^k})\right\}\\
B^Z_1&=&\frac{1}{16\pi^2}\lambda^{\prime}_{i2k}\lambda^{\prime}_{i3k}
\left \{ (v_e+a_e)M_t \left [ c_{11}-c_{12}+c_{21}-c_{23}\right ]
                  (-p_t,p_c,M_{e^i},M_{\tilde d^k},M_{e^i}) \right.\nonumber\\
& &\left. + \xi_V M_t \left[c_{11}-c_{12}+c_{21}-c_{23}\right]
                  (-p_t,k,M_{e^i},M_{\tilde d^k},M_{\tilde d^k})\right\}\\
A^Z_2&=&\frac{1}{16\pi^2}\lambda^{\prime}_{i2k}\lambda^{\prime}_{i3k}
\left \{ (v_c+a_c)B_1(M_t,M_{d^k},M_{\tilde e^i}) \right.\nonumber\\
& & - (a_d-v_d)\left[2c_{24}-\frac{1}{2}+M_Z^2(c_{12}+c_{23})\right]
                  (-p_t,p_c,M_{d^k},M_{\tilde e^i},M_{d^k})\nonumber\\
& &\left. - \xi'_V \left[ 2c_{24}+M_t^2 (c_{11}-c_{12}+c_{21}-c_{23})\right]
                  (-p_t,k,M_{d^k},M_{\tilde e^i},M_{\tilde e^i})\right\}\\
B^Z_2&=&\frac{1}{16\pi^2}\lambda^{\prime}_{i2k}\lambda^{\prime}_{i3k}
\left\{ (a_d-v_d)M_t\left[c_{11}-c_{12}+c_{21}-c_{23}\right]
                  (-p_t,p_c,M_{d^k},M_{\tilde e^i},M_{d^k})\right.\nonumber\\
& &\left. - \xi'_V M_t\left[c_{11}-c_{12}+c_{21}-c_{23}\right]
                  (-p_t,k,M_{d^k},M_{\tilde e^i},M_{\tilde e^i})\right\}.
\end{eqnarray}
Sum over family indices, $i, k=1,2,3$, is implied.
$p_t$ and $p_c$ are the momenta of the top and the charm quarks.
The functions $B_1$ and $c_{ij}$ are 2- and 3-point Feynman integrals given 
in [22], and their functional dependences are indicated in the bracket 
following them.  The constant 
$\xi_V(\xi'_V)= -e_d s_W/c_W\left(-(1-2s_W^2)/2 s_W c_W\right ),~
e_d(-1),~1(0)$ are for the $Z$ boson, photon and gluon, respectively; 
$v_f=(I^f_3-2e_fs_W^2)/2s_Wc_W$ and $a_f=I^f_3/2s_Wc_W$ are
the vector and axial-vector couplings 
with $e_f$ being the electric charge of the fermion $f$ in unit of $e$, 
and $I_3^f=\pm 1/2$ the corresponding third components of the weak isospin. 
The form factors $B^{\gamma}_{1,2}$ and $B^g_{1,2}$ 
are obtained from $B^Z_{1,2}$ by the substitutions,
$B^{\gamma}_1=B^Z_1 (a_e\rightarrow 0, v_e\rightarrow e_e)$,
$~B^{\gamma}_2=B^Z_2(a_d\rightarrow 0, v_d\rightarrow e_d)$,
$~B^g_1=B^Z_1(a_e\rightarrow 0, v_e\rightarrow 0)$ and
$B^g_2=B^Z_2(a_d\rightarrow 0, v_d\rightarrow 1)$; and setting
$M_Z\rightarrow 0$.

Note that the ultraviolet divergencies are contained in Feynman integrals 
$B_1$ and $c_{24}$. We have checked that all the ultraviolet divergencies  
cancelled as a result of renormalizability of MSSM.
\begin{figure}[tb]
\begin{center}
\psfig{figure=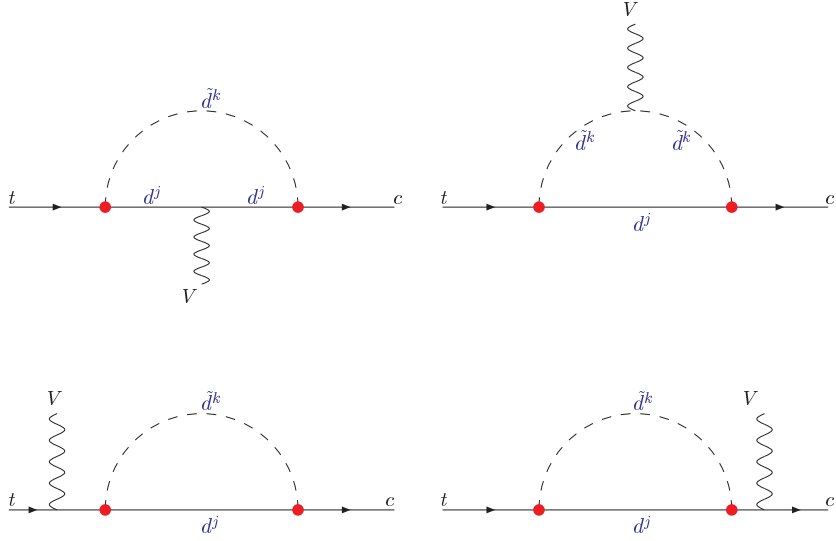,width=350pt,angle=0}
\end{center}
\vspace*{-0.5cm}
\caption{Feynman diagrams for $t\rightarrow cV $ ($V=Z,\gamma, g$
for quarks and squarks; $V=Z,\gamma$ for leptons and sleptons)
induced by B-violating couplings. The blobs denote B-violating vertex.}
\label{fig2}
\end{figure}

Similarly, we have calculated the effective $tcV$ vertices induced by the 
${\large \not} \! B$ couplings shown in Fig.2. 
The effective vertices have the forms similar to those of 
Eqs.(\ref{verz}-\ref{verg}) with the substitutions 
$P_L\leftrightarrow P_R$, $A^V\rightarrow F_1^V$ and $B^V\rightarrow F_2^V$, 
where 
\begin{eqnarray}
F^Z_1&=&\frac{1}{8\pi^2}
       \lambda^{\prime\prime}_{2jk}\lambda^{\prime\prime}_{3jk}
\left \{ (v_c-a_c)B_1(M_t,M_{d^j},M_{\tilde d^k})\right.\nonumber\\
& & + (v_d-a_d)\left[\frac{1}{2}-2c_{24}-M_V^2(c_{12}+c_{23})\right]
                  (-p_t,p_c,M_{d^j},M_{\tilde d^k},M_{d^j})\nonumber\\
& &\left. - \xi_V\left[2c_{24}+M_t^2(c_{11}-c_{12}+c_{21}-c_{23})\right]
                  (-p_t,k,M_{d^j},M_{\tilde d^k},M_{\tilde d^k})\right\}\\
F^Z_2&=&\frac{1}{8\pi^2}
 \lambda^{\prime\prime}_{2jk} \lambda^{\prime\prime}_{3jk}
\left \{ (v_d-a_d)M_t \left [c_{11}-c_{12}+c_{21}-c_{23}\right ]
                  (-p_t,p_c,M_{d^j},M_{\tilde d^k},M_{d^j})\right.\nonumber\\
 & &\left. - \xi_VM_t \left[c_{11}-c_{12}+c_{21}-c_{23}\right]
                  (-p_t,k,M_{d^j},M_{\tilde d^k},M_{\tilde d^k})\right\}\\
F^{\gamma}_2&=&F^Z_2\left\vert_{a_d\rightarrow 0, v_d\rightarrow e_d},~~
F^g_2=\frac{1}{2}F^Z_2\left\vert_{a_d\rightarrow 0, v_d\rightarrow -1,
                                 \xi_V\rightarrow -\xi_V}.
                                   \right. \right. 
\end{eqnarray}
Sum over family indices, $j, k=1,2,3$, is implied. 
\vspace{.5cm}

Now we present the numerical results for $Br(t\rightarrow cV)$.
We take $M_t=175$ GeV, $m_Z=91.187$ GeV, $m_W=80.3$ GeV, 
$G_F=1.16639\times 10^{-5}$(GeV)$^{-2}$, $\alpha=1/128$,
$\alpha_s=0.108$, and neglect the masses of charged leptons,
down-type quarks, and the charm quark.  The decay rates increase with 
the relevant $\lambda^{\prime}$ or $\lambda^{\prime\prime}$  couplings 
and decrease with the increase of the sparticle mass.

We note that there are two mass eigenstates for each flavor squark and slepton,
and the non-zero off-diagonal terms in the sfermion mass matrix will induce 
the mass splitting between the two mass eigenstates [23].
Since the off-diagonal terms in the mass matrix 
are proportional to the mass of the corresponding fermion [23], 
the off-diagonal terms in the mass matrix of the down-type squark 
and the slepton are relatively small. For simplicity, we assumed
all the down-type squark masses to be degenerate, so are the
mass of the sleptons. As we shall discuss later, these technical 
assumptions do not affect our results.   

{\bf L-violating Couplings}:~~
To calculate the bounds of the $Br(t\rightarrow c V)$ in the presence of
the ${\large \not} \! L$ terms, we use the following limits 
 on the ${\large \not} \! L$ couplings (obtained for the squark mass of 100 GeV): $
\vert \lambda^{\prime}_{kij}\vert<0.012~(k,j=1,2,3; i=2)$ [16],
$~\vert \lambda^{\prime}_{13j}\vert<0.16~(j=1,2)$ [18], 
$~\vert \lambda^{\prime}_{133}\vert<0.001$ [15],
$~\vert \lambda^{\prime}_{23j}\vert<0.16~(j=1,2,3)$
and $\vert \lambda^{\prime}_{33j}\vert<0.26~(j=1,2,3)$ [19].
There are also the following constraints on the products
of the $\lambda^{\prime}$ couplings [17][18]:
$\lambda^{\prime}_{13i}\lambda^{\prime}_{12i},~
\lambda^{\prime}_{23j}\lambda^{\prime}_{22j}<1.1\times 10^{-3} 
~(i=1,2; j=1,2,3)$,
$~\lambda^{\prime}_{in2}\lambda^{\prime}_{jn1} <10^{-5}~(i,j,n=1,2,3)$,
and $\lambda^{\prime}_{121}\lambda^{\prime}_{222},
~\lambda^{\prime}_{122}\lambda^{\prime}_{221},
~\lambda^{\prime}_{131}\lambda^{\prime}_{232},
~\lambda^{\prime}_{132}\lambda^{\prime}_{231}<10^{-7}$.

Using the upper limits of the relevant ${\large \not} \! L$ couplings
and taking the lower limit of 45 GeV for slepton mass, 
we find the maximum values of the branching fractions to be
\begin{eqnarray}
Br(t \rightarrow cZ) \leq 10^{-9}, 
~Br(t \rightarrow c\gamma)\leq  10^{-10},
~Br(t \rightarrow cg)     \leq  10^{-8}.
\end{eqnarray}
If we consider the mass splitting between sleptons, 
these upper limits on the branching fractions still persist.
Thus we conclude that the contributions of the ${\large \not} \! L$ 
couplings to $t \rightarrow cV$ are too small to be of interest. 
\vspace{.5cm}

{\bf B-violating Couplings}:~~
For the ${\large \not} \! B$ couplings, $\lambda^{\prime\prime}$,
the bound on top rare decay rates can be significantly increased 
since the  $\lambda^{\prime\prime}$ couplings stand relatively unconstrained, 
except for $\lambda^{\prime\prime}_{112}$ and $\lambda^{\prime\prime}_{113}$
which  have been strongly bounded from the consideration 
of double nucleon decay into two kaons [12] and $n-\bar n$ oscillation [12],
respectively. 

Under the assumption that the masses of all down-type squarks
are degenerate, $Br(t\rightarrow c V)$ is proportional to $\Lambda^2$
with $\Lambda$ being the product of the relevant
${\large \not} \! B$ couplings defined by
\begin{equation}\label{lam}
\Lambda\equiv  
 \lambda^{\prime\prime}_{212} \lambda^{\prime\prime}_{312}
                +\lambda^{\prime\prime}_{213} \lambda^{\prime\prime}_{313}
                +\lambda^{\prime\prime}_{223} \lambda^{\prime\prime}_{323}
          =\frac{1}{2}\lambda^{\prime\prime}_{2jk} \lambda^{\prime\prime}_{3jk}.
\end{equation}
While the experimental bounds on $\lambda^{\prime\prime}_{3jk}$ have been 
derived from the ratio of hadron to lepton width of the $Z^0$, 
$R_l\equiv \Gamma_h/\Gamma_l$ [20], we are not aware of any experimental 
bounds on $\lambda^{\prime\prime}_{2jk}$ although one can make general 
estimates from certain low energy data. Therefore, we do not have an 
experimental bound for $\Lambda$. We discuss these points in some detail below.

First we will argue that it is likely only one term in $\Lambda$, 
Eq.(\ref{lam}), can be significant. This comes from the consideration of 
the low energy processes, $b\rightarrow s\gamma$ and $K^0-\bar K^0$ mixing.
They may provide strong constraints to the products 
$\lambda^{\prime\prime}_{i12} \lambda^{\prime\prime}_{i13}$
and $\lambda^{\prime\prime}_{i13} \lambda^{\prime\prime}_{i23}$
(sum over $i$ is implied), respectively  [24]. 
Thus the simultaneous presence of any two terms in
$\Lambda$ might conflict with these low energy processes.
However, the existence of only one term,
 $\lambda^{\prime\prime}_{212} \lambda^{\prime\prime}_{312}$,
             $\lambda^{\prime\prime}_{213} \lambda^{\prime\prime}_{313}$ or
             $\lambda^{\prime\prime}_{223} \lambda^{\prime\prime}_{323}$,
will not be  constrained by them.

The bound on $\lambda^{\prime\prime}_{3jk}$ from 
$R_l\equiv \Gamma_h/\Gamma_l$ is 1.46 at $2\sigma$ for down squark mass 
of 100 GeV [20]. 
We can obtain another constraint from the FNAL data of $t\bar t$ 
events by examing the exotic top quark decay
$t\rightarrow \bar d_L^j+\bar {\tilde d_R^k}$.
For the top mass of 175 GeV, we have
\begin{equation}
R_t\equiv\frac{\Gamma(t\rightarrow \bar d_L^j+\bar {\tilde d_R^k})}
{\Gamma(t\rightarrow W+b)}=1.12\left (\lambda''_{3jk}\right )^2\left [ 1-
 \left(\frac{M_{\tilde d^k_R}}{175 {\rm GeV}}\right )^2\right ]^2
\theta(1-\frac{m_{\tilde d}}{m_t}).
\end{equation}
The ${\tilde d_R^k}$ can decay into a $d_R$ plus a lightest neutralino
(and gluino if kinematically allowed), 
as well as quark pairs induced by the ${\large \not} \! B$ terms.
 The decay modes, $t\rightarrow \bar d_L^j+\bar {\tilde d_R^k}$
can enhance the total fraction of
hadronic decays of the top quark and alter the ratio of $t \bar t$ 
events expected in the dilepton channel.
The number of dilepton events expected in the presence of the decay
$t\rightarrow \bar d_L^j+\bar {\tilde d_R^k}$ and that
in the SM is given by $R(f)\equiv (1-f)^2$, where 
f = $Br (t\rightarrow \bar d_L^j+\bar {\tilde d_R^k} ) $.
The CDF measurements of the $t\bar t$ production cross section is
$\sigma[t\bar t]_{\rm exp}=8.3^{+4.3}_{-3.3}$ pb in the dilepton channel[25],
while the SM expectation for top mass of 175 GeV is
$\sigma[t\bar t]_{\rm QCD}=5.5^{+0.1}_{-0.4}$ pb [26]. By requiring 
$R(f)$ to lie within the measured range of $\sigma[t\bar t]_{\rm exp}/ 
\sigma[t\bar t]_{\rm QCD}$, we can obtain the bounds on the relevant 
$\lambda''$ couplings. The $2\sigma$ bound from dilepton channel is 
found to be 
\begin{equation}
 \left (\lambda''_{3jk}\right )^2\left [ 1-
 \left(\frac{M_{\tilde d^k_R}}{175 {\rm GeV}}\right )^2\right ]^2<0.71.
\end{equation}
For $M_{\tilde d^k_R}= 100$ GeV, we have $\lambda_{3jk}^{\prime\prime}<1.25$,
comparable to the bound from $R_l$ [20]. Constraints on
 $\lambda_{3jk}^{\prime\prime}$ from the experimental data of $t \bar t$
in other channels are weaker.

 Although we are not aware of any experimental bound for 
$\lambda^{\prime\prime}_{2jk}$, theoretical bounds can be derived 
under specific assumptions [11].  The constraint of perturbative 
unitarity at the SUSY breaking scale $M_{SUSY}$  would bound all the couplings,
and in particular $\frac{(\lambda^{\prime\prime}_{2jk})^2}{4\pi}<1$, i.e., 
$\lambda^{\prime\prime}_{2jk}<3.54$.  A stronger bound can be obtained if we 
assume the gauge group unification at $M_U=2\times 10^{16}$ GeV and the 
Yukawa couplings $Y_t, Y_b$ and $Y_{\tau}$ to remain in the perturbative 
domain in the whole range up to $M_U$. They imply $Y_i(\mu)<1$ for 
$\mu<2\times 10^{16}$ GeV. Then we obtain an upper bound of 1.25 
for all $\lambda^{\prime\prime}_{ijk}$ [11]. In this latter case, 
in case of the presence of one term (three terms) in $\Lambda$,
$\Lambda$ has its maximum value of 1.6 (4.7). 
But there is no a priori reason to take this theoretical
assumption. Taking the former scenario of perturbative unitarity at
the SUSY breaking scale and let, for example, 
$\lambda^{\prime\prime}_{212}$ and  $\lambda^{\prime\prime}_{312}$
having their maximal allowed values and all the other 
$\lambda^{\prime\prime}$'s  to be small, then we have  
$\Lambda$ as large as 5.
 
Now we present the numerical results for the effects of  
$\lambda^{\prime\prime}$ couplings by considering $\Lambda$ as 
a variable and dividing it out from the branching ratios.

\begin{figure}[tb]
\begin{center}
\psfig{figure=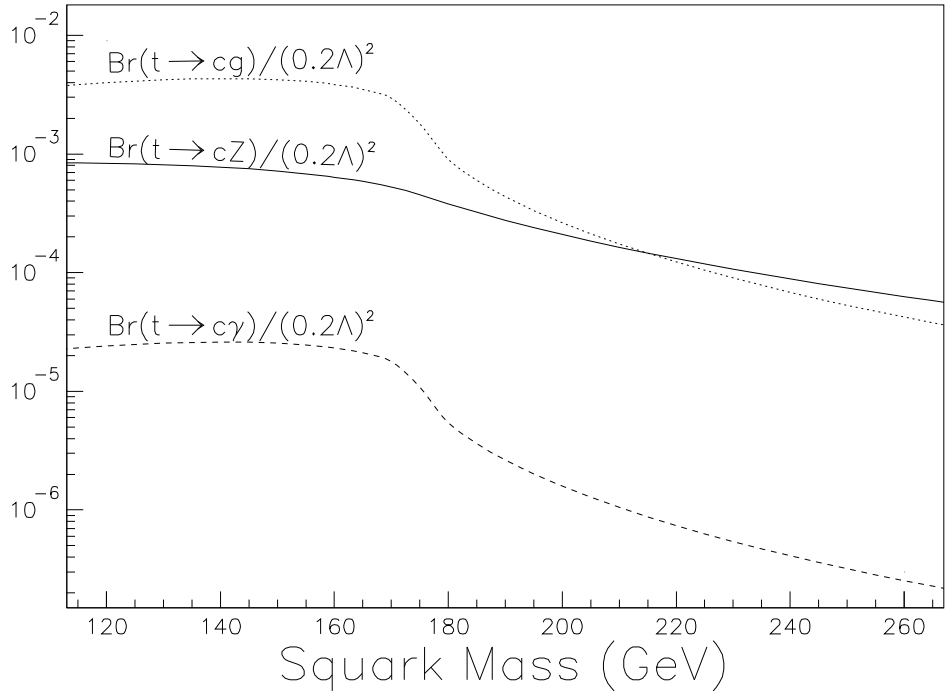,width=300pt,angle=-90}
\end{center}
\vspace*{-1.3cm}
\caption{The plot of $Br(t \rightarrow cV)/(0.2\Lambda)^2$ 
  as a function of squark mass.}
\label{fig3}
\end{figure}
In Fig.3 we present the plot of $Br(t \rightarrow cV)/(0.2\Lambda)^2$ 
as a function of squark mass. For squark mass no greater than 170 GeV we have  
\begin{eqnarray}\label{ratio1}
Br(t \rightarrow cZ)      \approx (1.2\Lambda)^2\times 10^{-4},\\
\label{ratio2}  
Br(t \rightarrow c\gamma)\approx (0.6\Lambda)^2\times 10^{-5},\\
\label{ratio3}
Br(t \rightarrow cg)     \approx (0.4\Lambda)^2\times 10^{-3}.
\end{eqnarray}
We conclude from eqs.(\ref{ratio1}-\ref{ratio3}),
eqs.(\ref{level1})-(\ref{level5}) and $\Lambda$ to be as large as 5
that the contribution of ${\large \not} \! B$ couplings to the decay 
$t \rightarrow cV$ might be observable at the upgraded Tevatron and LHC. 

If the decays $t \rightarrow cV$ are not observed at the upgraded Tevatron 
and LHC, we can obtain the experimental upper bound for $\Lambda$.
We illustrate this in Fig.4 where we plot $\Lambda$ 
versus the degenerate squark mass.
The solid, dashed and dotted lines correspond to 
$Br(t \rightarrow cg)=1\times 10^{-3}$,
$Br(t \rightarrow cZ)=2\times 10^{-4}$ and 
$Br(t \rightarrow c\gamma)=5\times 10^{-6}$, respectively.
The region above the solid line corresponding to 
$Br(t \rightarrow cg)>1\times 10^{-3}$ will be excluded
if the decay  $t \rightarrow cg$ is not observed 
at the upgraded Tevatron.
The region above the dashed and dotted line corresponds to 
$Br(t \rightarrow cZ)>2\times 10^{-4}$ and 
$Br(t \rightarrow c\gamma)>5\times 10^{-6}$ which
will be excluded if corresponding decays
are not observed  at the LHC.  The corresponding value of $\Lambda$
which sets its upper bound  can be read off from the figure. 
For example, for squark mass of 150 GeV, 
the upgraded Tevatron can probe the $\Lambda$ down to 2.3.
This bound is not very strong but may serve as the first 
experimental bound on this hitherto experimentally unconstrained product 
of $\lambda''$ couplings.
\begin{figure}[tb]
\begin{center}
\psfig{figure=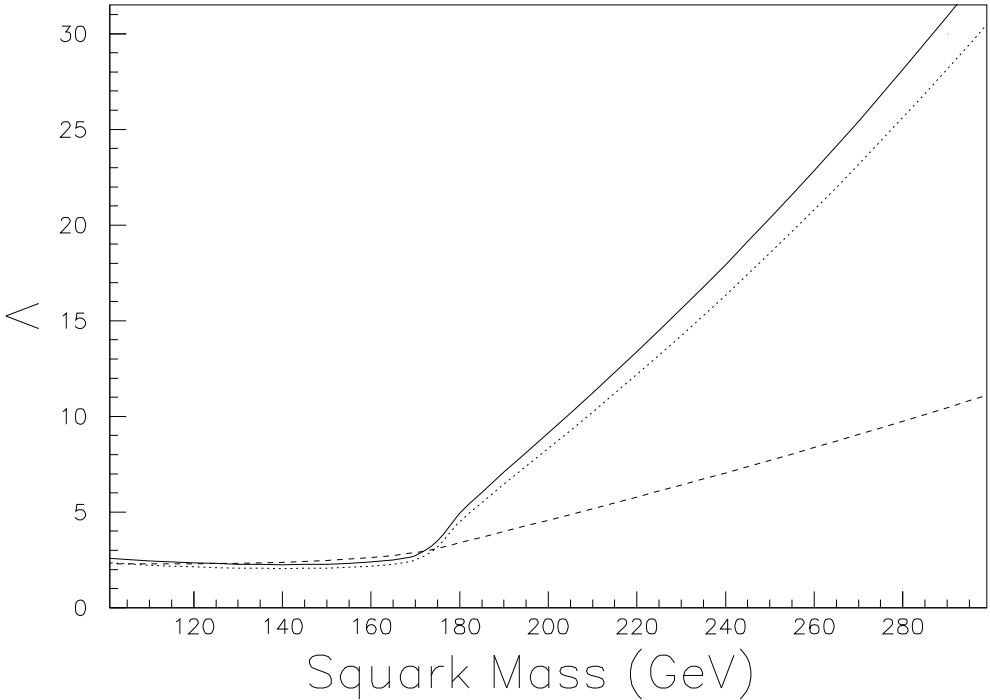,width=300pt,angle=-90}
\end{center}
\vspace*{-1.3cm}
\caption{ $\Lambda$ versus squark mass for given values of branching ratios.
The solid, dashed and dotted lines correspond to 
$Br(t \rightarrow cg)=1\times 10^{-3}$,
$Br(t \rightarrow cZ)=2\times 10^{-4}$ and 
$Br(t \rightarrow c\gamma)=5\times 10^{-6}$, respectively.}
\label{fig4}
\end{figure}
A few remarks are due regarding the above results:
\begin{itemize}
\item[{\rm(a)}]
    For the upgraded Tevatron or LHC, the limits on some individual 
    or combinations of these couplings may be obtainable from direct 
    squark search. However, we think that our results are complementary  
    to the direct search and the processes discussed in  the present article 
    may involve different combination of the couplings. Since the R-violating 
    SUSY contains many parameters, it is desirable to obtain as many 
    constraints as possible.
\item[{\rm(b)}]
    If the HERA anomalous events [27] were the result of R-parity violating 
    terms [28], namely non-zero values for ${\large \not} \! L$ couplings 
    $\lambda'$, all the ${\large \not} \! B$ couplings  
    $\lambda^{\prime\prime}$ would be very small since proton stability 
    imposes a upper bound of $10^{-9}(10^{-11})$ for any products of 
    $\lambda^{\prime}\lambda^{\prime\prime}$ in the absence (presence) of 
    squark flavor mixing [14]. Then the effects of any 
    $\lambda^{\prime\prime}$ coupling would of course not be observable.
\item[{\rm(c)}]
   As we have pointed out, only one term in $\Lambda$ can exist, which is
   $\lambda^{\prime\prime}_{212} \lambda^{\prime\prime}_{312}$,
   $\lambda^{\prime\prime}_{213} \lambda^{\prime\prime}_{313}$ or
   $\lambda^{\prime\prime}_{223} \lambda^{\prime\prime}_{323}$.
  Let us assume the existence of  
  $\lambda^{\prime\prime}_{212} \lambda^{\prime\prime}_{312}$ as an example.
  Besides the two-body rare decays $t\rightarrow cV$, 
  the three-body decays $t\rightarrow c d\bar d$ (exchanging
  a $\tilde s$) and  $t\rightarrow c s\bar s$ (exchanging a $\tilde d$)
  can aslo open. Although these decay modes just give rise to three light
  jets and thus are not easy to detect at the upgraded Tevatron or LHC, 
  detailed examination for the possibility of detecting these decay modes 
  are needed [24].
\item[{\rm(d)}]
    In the contributions of $\lambda^{\prime\prime}$ couplings, 
    the masses of down-type squarks $\tilde d$, $\tilde s$ and $\tilde b$
    are involved. In our calculation, we assumed the degeneracy of these
    masses so that we extracted a factor $\Lambda$ in Eq.(\ref{lam}).
    However, as  we have pointed out, only one term in $\Lambda$ can exist.
    Correspondingly, only one flavor of down-type squark is involved.
    So actually our assumption of mass degeneracy between different 
    flavor down-type squarks does not affect our numerical results.
  
    Further, we assumed the mass degeneracy between the two
    mass eigenstates for each flavor down-type squark. 
    Again, let us assume the existence of the term   
    $\lambda^{\prime\prime}_{212} \lambda^{\prime\prime}_{312}$ as an example.
    Then only strange-squark ($\tilde s$) is involved. 
    There are two mass eigenstates for it, 
    namely $\tilde s_1$ and $\tilde s_2$.
    We assumed  $m_{\tilde s_1}=m_{\tilde s_2}$ in our calculation.
    Theoretically this is a good approximation because the mass splitting 
    between $\tilde s_1$ and $\tilde s_2$ is
    proportional to the strange-quark mass [23]. 
    We also checked that our numerical results are not sensitive
    to the small mass-splitting.

    Althogh the possible mass-splittings between different flavor squarks 
    cannot significantly enhance the rates of top rare decays, they would 
    cause some unexpected effects in low energy processes. For example, 
    the large mass-splitting between charm-squark and up-squark, which 
    are not relevant to our calculations in this paper,
    would lead to large FCNC processes in the D-meson system.
    This will be examined in detail in our future work.         
\item[{\rm(e)}]  
    Finally, we should point out that 
    with some couplings as large as 3.5, the model
    cannot be extrapolated beyond a few TeV, which 
    will take away many of the motivations of supersymmetry.
\end{itemize}
\vspace{.5cm}

In summary, we found that the decays $t \rightarrow cV$ can be significantly 
enhanced relative to those in the R-parity conserving SUSY model. 
In an optimistic scenario that one of the products in $\Lambda$ of 
Eq.(\ref{lam}) attend the allowed limit by perturbative unitarity at the 
SUSY breaking scale and by $R_l$,  the branching ratios can be as large as
$Br(t \rightarrow cg) \sim 10^{-3}$,
$Br(t \rightarrow cZ) \sim 10^{-4}$ and $Br(t \rightarrow c\gamma)
\sim 10^{-5}$, which are potentially observable at the upgraded Tevatron and/or
the LHC. If not seen, upper bounds can be set on the specific combination
of the relevant ${\large \not} \! B$ couplings.
Together with low energy processes such as $b\rightarrow s\gamma$ and 
$K^0-\bar K^0$ mixing, strong bounds on most of the  
$\lambda^{\prime\prime}$ couplings can be set.
\vspace{.5cm}

We thank  A. Datta, T. Han and C. Kao for discussions. This work 
was supported in part by the U.S. Department of Energy, Division of 
High Energy Physics, under Grant No. DE-FG02-94ER40817 and DE-FG02-92ER40730.
XZ was also supported in part by National Natural Science Foundation of China
and JMY acknowledge the partial support provided by the Henan Distinguished
Young Scholars Fund.
\vspace{.5cm}

{\LARGE References}
\vspace{0.3in}
\begin{itemize}
\item[{\rm[1]}] J. L. Diaz-Cruz et al., Phys. Rev. D{\bf 41}, 891 (1990);
             B. Dutta Roy et al., Phys. Rev. Lett.65, 827 (1990); 
             H. Fritzsch, Phys. Lett. B{\bf 224}, 423 (1989); 
             W. Buchmuller and M. Gronau, Phys. Lett. B{\bf 220} 641 (1989).
\item[{\rm[2]}]  G. Eilam, J. L. Hewett and A. Soni, Phys. Rev. D{\bf 44}, 
                 1473 (1991).
\item[{\rm[3]}] J. Incandela (CDF), FERMILAB-CONF-95/237-E (1995) (unpublished);
                D. Gerdes, hep-ex/9706001, 
                to appear in the Proceedings of the XXXIIth
                Rencontres de Moriond, Electroweak Interactions and 
                Unified Theories, Les Arcs, Savoie, France, March 15-22, 1997
\item[{\rm[4]}] T. J. Lecompte (CDF), FERMILAB-CONF-96/021-E (1996) (unpublished.
\item[{\rm[5]}] A. P. Heinson (D0), hep-ex/9605010 (unpublished).
\item[{\rm[6]}] T. Han, R. D. Peccei, and X. Zhang, Nucl. Phys. {\bf B454},
                527 (1995);  
                T. Han, K. Whisnant, B.-L. Young, and X. Zhang,
                Phys. Rev. D55, 7241 (1997);
                T. Han, K. Whisnant, B.-L. Young, and X. Zhang, 
                Phys. Lett. B{\bf 385}, 311 (1996); 
                A. Heison, FERMILAB-CONF-96/116-E, May 1996.
\item[{\rm[7]}] T. Tait and C.-P. Yuan,  Phys. Rev. D55,  7300 (1997);
                E. Malkawi and T. Tait,  Phys. Rev. D54,  5758 (1996); 
                M. Hosch, K. Whisnant, and  B.-L. Young, Phys.Rev. D56, 5725
                (1997);
                ``Future Electroweak Physics at the Fermilab Tevatron",
                 Report of the TeV 2000 working group, FERMILAB-PUB-96/082,
                 edited by D. Amidei and R. Brock.
\item[{\rm[8]}] C. S. Li, R. J. Oakes and J. M. Yang, Phys. Rev. D{\bf 49}, 293 (1994); 
                J. M. Yang and C. S. Li, Phys. Rev. D{\bf 49}, 3412 (1994); 
                G. Couture, C. Hamzaoui and H. Konig, Phys. Rev. D{\bf 52}, 1713 (1995); 
              J. L. Lopez, D. V. Nanopoulos and R. Rangarajan, 
              Phys. Rev. D56, 3100  (1997);
                G. Couture, M. Frank and H. Konig, Phys. Rev. D56, 4213 (1997);
                G. M. de Divitiis, R. Petronzio and L. Silvestrini,
                                                Nucl. Phys. B504, 45 (1997).
\item[{\rm[9]}] For reviews of the MSSM, see, for example,
                 H. E. Haber and G. L. Kane, Phys. Rep. {\bf 117}, 75  (1985); 
                 J. F. Gunion and H. E. Haber, Nucl. Phys. {\bf B272}, 1  (1986). 
\item[{\rm[10]}] For reviews of R-parity-violating couplings, 
                  see, for example,
                 D. P. Roy, hep-ph/9303324 (unpublished) ; 
                 G. Bhattacharyya, Nucl. Phys. Proc. Suppl. 52A, 83  (1997).
\item[{\rm[11]}] B. Brahmachari and P. Roy, Phys. Rev. D{\bf 50}, 39 (1994).
\item[{\rm[12]}] J. L. Goity and M. Sher, Phys. Lett. B {\bf 346}, 69 (1995);
\item[{\rm[13]}]
               R. N. Mohapatra, Phys. Rev. D{\bf 34}, 3457 (1986); 
                M. Hirsch, H. V. Klapdor-Kleingrothaus, S. G. Kovalenko,
                Phys. Rev. Lett. {\bf 75}, 17 (1995); 
                K. S. Babu and R. N. Mohapatra, Phys. Rev. Lett. {\bf 75}, 
                2276 (1995);
                 V. Barger, G. F. Giudice and T. Han, Phys. Rev. D{\bf 40}, 
                 2978 (1989); 
                 G. Bhattacharyya and D. Choudhury, Mod. Phys. Lett. A{\bf 10}, 
                 1699 (1995);
                 D. E. Kaplan, hep-ph/9703347 (unpublished); 
                   J. Jang, J. K. Kim and J. S. Lee, 
                  Phys. Rev. D55, 7296 (1997).
\item[{\rm[14]}] A.Y.Smirnov and F.Vissani, Phys. Lett. B{\bf 380}, 317 (1996);
                 F. Zwirner,  Phys. Lett. B{\bf 132}, 103 (1983). 
\item[{\rm[15]}] S. Dimopoulos and L. J. Hall, Phys. Lett. B{\bf 207}, 210 (1987); 
                R. M. Godbole, P. Roy and X. Tata, Nucl. Phys. {\bf B401}, 67 (1993). 
\item[{\rm[16]}] K. Agashe and M. Graesser, Phys. Rev. D{\bf 54}, 4445 (1996). 
\item[{\rm[17]}] D. Choudhury and P. Roy, Phys. Lett. B378,  153 (1996).
\item[{\rm[18]}] J. Jang, J. K. Kim and J. S. Lee, Phys. Lett. B408, 367 (1997).
\item[{\rm[19]}] G. Bhattacharyya, J. Ellis and K. Sridhar,
                  Mod. Phys. Lett. A{\bf 10},1583 (1995). 
\item[{\rm[20]}] G. Bhattacharyya, D. Choudhury and K. Sridhar, 
                 Phys. Lett. B{\bf 355}, 193  (1995). 
\item[{\rm[21]}] J. Erler, J. L. Feng and N. Polonsky, Phys. Rev. Lett. 78, 
                 3063 (1997); 
                 D. K. Ghosh, S. Raychaudhuri and K. Sridhar,  
                 Phys. Lett. B396,  177 (1997);
                 D. Choudhury and S. Raychaudhuri,  Phys. Lett. B401, 54 (1997);
                 A. Datta, J. M. Yang, B.-L. Young and X. Zhang, 
                 Phys. Rev. D56, 3107 (1997).
\item[{\rm [22]}] G. Passarino and M. Veltman, Nucl. Phys. {\bf B160}, 
                  151 (1979).
\item[{\rm [23]}] J. Ellis and S. Rudaz, Phys. Lett. {\bf B128}, 248 (1983);
                  A. Bouquet, J. Kaplan and C. Savoy, Nucl. Phys. B262,
                  299 (1985). 
\item[{\rm [24]}] J. M. Yang et al., work in progress.
\item[{\rm [25]}] D. Gerdes, hep-ex/9609013 (unpublished).
\item[{\rm [26]}] See, for example, E. L. Berger and H. Contopanagos,
                 Phys. Lett. B {\bf 361}, 115 (1995). 
\item[{\rm[27]}] H1 Collab., C. Adloff et al., DESY 97-024; 
                 ZEUS Collab., J. Breitweg et al., DESY 97-025. 
\item[{\rm[28]}]  D. Choudhury and S. Raychaudhuri, Phys. Lett. B401, 
                  54 (1997); 
                  G. Altarelli, J. Ellis, G. F. Guidice, S. Lola and M. L. 
                  Mangano, Nucl. Phys. B506,  3 (1997); 
                  H. Dreiner and P. Morawitz, Nucl. Phys. B503,  55 (1997); 
                  J. Kalinowski, R. R\"uckl, H. Spiesberger and  P. M. Zerwas,
                  Z. Phys. C74, 595  (1997). 
\end{itemize}
\end{document}